\documentclass[journal]{IEEEtran}
\usepackage{amsthm}
\usepackage{amssymb}
\usepackage{graphicx}
\usepackage{amsmath}
\usepackage{bm}
{}
{}
\newtheorem{lemma}{Lemma}{}
{}
{}

\usepackage{algorithm}
\usepackage{algpseudocode}
\usepackage{graphics}
\usepackage{epsfig}
\usepackage{multirow}
\usepackage{array}
\usepackage{cite}
\usepackage{stfloats}
\usepackage{subfigure}
\usepackage{booktabs}
\usepackage{color}
\usepackage{tabu} 
\usepackage{stfloats}
\usepackage{mathtools}
\usepackage{stmaryrd} 

\begin{document}
\title{Channel Estimation for Movable-Antenna MIMO Systems Via Tensor Decomposition}


\author{\IEEEauthorblockN{
		Ruoyu Zhang, \IEEEmembership{Member,~IEEE}, 
		Lei Cheng, \IEEEmembership{Member,~IEEE}, 
		Wei Zhang, \IEEEmembership{Member,~IEEE}, \\
		Xinrong Guan, 
		Yueming Cai,
		Wen Wu, \IEEEmembership{Senior Member,~IEEE}, 
		and Rui Zhang, \IEEEmembership{Fellow,~IEEE} 
	}
\thanks{
	This work was supported in part by the National Natural Science Foundation of China under Grants 62201266, 62371418, 62171461, in part by the Natural Science Foundation of Jiangsu Province under Grant BK20210335, and in part by the Shenzhen Science and Technology R\&D Funds under Grant No. GXWD20231130104830003.
	\emph{(Corresponding author: Xinrong Guan).} }
\thanks{
	R. Zhang and W. Wu are with the Key Laboratory of Near-Range RF Sensing ICs \& Microsystems (NJUST), Ministry of Education, School of Electronic and Optical Engineering, Nanjing University of Science and Technology, Nanjing 210094, China (e-mail: ryzhang19@njust.edu.cn; wuwen@njust.edu.cn).
	L. Cheng is with the College of Information Science and Electronic Engineering, Zhejiang University, Hangzhou 310027, China (e-mail: lei\_cheng@zju.edu.cn).
	W. Zhang is with the School of Electronics and Information Engineering, Harbin Institute of Technology Shenzhen, 518055, China (e-mail: zhangwei.sz@hit.edu.cn).
	X. Guan and Y. Cai are with the College of Communications Engineering, Army Engineering University of PLA, Nanjing 210007, China (e-mail: guanxr@aliyun.com; caiym@vip.sina.com).
	Rui Zhang is with the School of Science and Engineering, Shenzhen Research Institute of Big Data, The Chinese University of Hong Kong, Shenzhen, Guangdong 518172, China, and also with the Department of Electrical and Computer Engineering, National University of Singapore, Singapore 117583 (e-mail: rzhang@cuhk.edu.cn; elezhang@nus.edu.sg). }
\vspace{-1.75em}
}


\maketitle



%
\IEEEpeerreviewmaketitle

\begin{abstract}
In this letter, we investigate the channel estimation problem for MIMO wireless communication systems with movable antennas (MAs) at both the transmitter (Tx) and receiver (Rx).
To achieve high channel estimation accuracy with low pilot training overhead, we propose a tensor decomposition-based method for estimating the parameters of multi-path channel components, including their azimuth and elevation angles, as well as complex gain coefficients, thereby reconstructing the wireless channel between any pair of Tx and Rx MA positions in the Tx and Rx regions.
First, we introduce a two-stage Tx-Rx successive antenna movement pattern for pilot training, such that the received pilot signals in both stages can be expressed as a third-order tensor. 
Then, we obtain the factor matrices of the tensor via the canonical polyadic decomposition, and thereby estimate the angle/gain parameters for enabling the channel reconstruction between arbitrary Tx/Rx MA positions.
In addition, we analyze the uniqueness condition of the tensor decomposition, which ensures the complete channel reconstruction between the whole Tx and Rx regions based on the channel measurements at only a finite number of Tx/Rx MA positions.
Finally, simulation results are presented to evaluate the proposed tensor decomposition-based method as compared to existing methods, in terms of channel estimation accuracy and pilot overhead.
	
\end{abstract}


\section{Introduction}

By exploiting the spatial degrees of freedom (DoFs) with multiple antennas, multiple-input multiple-output (MIMO) technology has been significantly advanced over decades to enhance the performance of generations of wireless networks.
However, due to the fixed geometry of conventional antenna arrays, MIMO systems cannot fully exploit the wireless channel spatial variation/DoFs, even for massive MIMO with a large number of fixed-position antennas \cite{lu2014overview,Zhang2024TWC}.

To overcome this limitation, movable antennas (MAs)/fluid-antenna systems (FAS) have been proposed to enable the flexible local movement of antennas within a confined spatial region at the transmitter (Tx) and/or receiver (Rx), for achieving better channel conditions to improve the communication performance \cite{Zhu2023Movable_Mag, Wong2021Fluid}. 
In \cite{Zhu2023Modeling}, a field-response channel model was introduced for the MA-based communication systems, which efficiently represents the channel between any pair of Tx and Rx antenna positions in terms of multi-path channel components. Based on this channel model, the positions of MAs are jointly optimized at the Tx and/or Rx to maximize the communication system spectral/energy efficiency under various MIMO/multiuser system setups \cite{Zhu2023Movable,Ma2024MIMO}.
However, the optimization of MAs' positions generally  requires the knowledge of all channels between any pair of positions in the whole Tx/Rx antenna movement regions, which is thus a more challenging task as compared to conventional MIMO system channel estimation with fixed Tx/Rx antenna positions.

In \cite{Skouroumounis2023Fluid}, a low-complexity port selection scheme was proposed for FAS to perform a linear minimum mean-squared error (MMSE)-based channel estimation with a small number of antenna ports selected. 
In \cite{Ma2023Compressed} and \cite{Xiao2024Channel}, a compressed sensing (CS)-based channel estimation method was proposed for MA systems with a single MA at both the Tx and Rx, where the angular-domain field-response parameters including the angles of departure (AoDs), angles of arrival (AoAs), and complex gain coefficients of multi-path channel components are estimated. 
However, this method relies on discrete angle quantization for sparse representation of channels, which inevitably limits the channel estimation accuracy.
Moreover, how to efficiently reconstruct the channel between any Tx/Rx positions with multiple MAs in the MA-enabled MIMO system still remains unknown.

To tackle the above challenges, we propose in this letter a new tensor decomposition-based channel estimation/reconstruction method for MA-enabled MIMO systems.
Particularly, a two-stage Tx-Rx successive antenna movement pattern is introduced for pilot training, such that the received pilot signals in both stages can be expressed as a third-order tensor. 
Then, we obtain the factor matrices of the tensor via the canonical polyadic (CP) decomposition and estimate the multi-path channel components, thereby reconstructing the wireless channel between arbitrary Tx/Rx MA positions.
In addition, we analyze the uniqueness condition of the tensor decomposition, which ensures the accurate and efficient channel reconstruction based on the channel measurements at only a finite number of Tx/Rx MA positions.
Simulation results demonstrate the superiority of the proposed tensor decomposition-based method in terms of channel estimation accuracy and pilot overhead, as compared to existing methods.

Notations:
The operations of transpose, conjugate, conjugate transpose, and Moore-Penrose pseudo-inverse 
are denoted by $\left( \cdot \right)^{T}$, 
$\left( \cdot \right)^{*}$, $\left( \cdot \right)^{H}$, and $\left( \cdot \right)^{\dagger}$, respectively.
$\mathbf{I}_{M}$ is the identity matrix with dimension $M \times M$, 
and $\mathrm{j} = \sqrt{-1}$ is an imaginary symbol. 
$[\mathbf{a}]_{m}$ represents the $m$-th entry of a vector $\mathbf{a}$. 
$\odot$ and $\otimes$ denote the Khatri-Rao product and Kronecker product, respectively. 

\section{System and Channel Model}

\begin{figure}[!tp]
	\centering
	\includegraphics[width=3.4in]
	{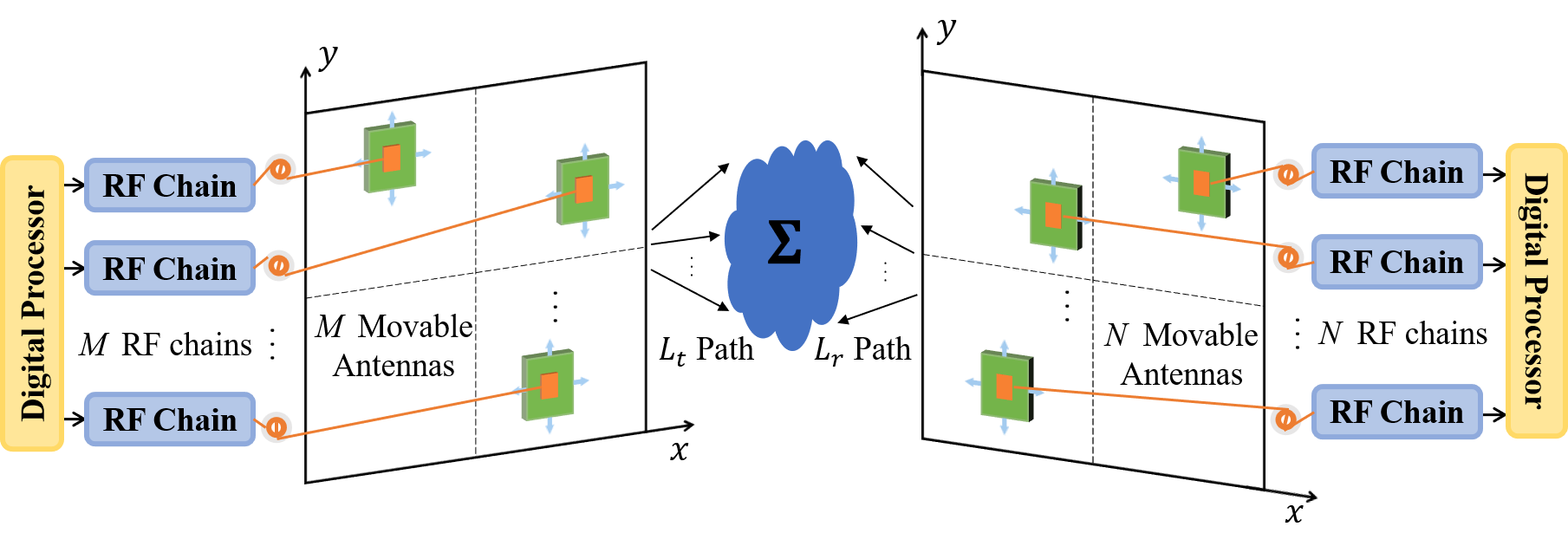}
	\caption{The MIMO system with MAs at both the transmitter and receiver.}
	\label{Fig_SystemModel}
\end{figure}

As shown in Fig. \ref{Fig_SystemModel}, we consider a MIMO system where the Tx and Rx are equipped with $M$ and $N$ MAs, respectively. 
The positions of transmitter-side MAs (T-MAs) and receiver-side MAs (R-MAs) can be flexibly adjusted over the two-dimensional local regions $\mathcal{C}^{t}$ and $\mathcal{C}^{r}$, which are assumed to be rectangular regions of size $A_{x} \times A_{y}$ and $B_{x} \times B_{y}$, respectively.  
For the $m$-th T-MA and $n$-th R-MA, ($m = 1, \ldots, M$, $n = 1, \ldots, N$), 
their positions can be represented by local Cartesian coordinates $\mathbf{t}_{m} = [x_{m}^{t}, y_{m}^{t}]^{T} \in \mathcal{C}^{t}$ and
$\mathbf{r}_{n} = [x_{n}^{r}, y_{n}^{r}]^{T} \in \mathcal{C}^{r}$, respectively.
Let $\tilde{\mathbf{t}} = [\mathbf{t}_{1}, \ldots, \mathbf{t}_{M}] \in \mathbb{R}^{2 \times M}$ and 
$\tilde{\mathbf{r}} = [\mathbf{r}_{1}, \ldots, \mathbf{r}_{N}] \in \mathbb{R}^{2 \times N}$ be the collections of coordinates of $M$ T-MAs and $N$ R-MAs, respectively.
Then, the input-output relation for the MA-enabled MIMO system is given by 
\begin{align}
\mathbf{y} = \sqrt {P} \mathbf{H}\big(\tilde{\mathbf{r}}, \tilde{\mathbf{t}}\big) \mathbf{s} + \mathbf{z}, \label{equ:input-output}
\end{align}
where $\mathbf{H} \left( \tilde{\mathbf{r}}, \tilde{\mathbf{t}} \right)  \in \mathbb{C}^{N \times M}$ is the MIMO channel matrix depending on the positions of T-MAs $\tilde{\mathbf{t}}$ and R-MAs $\tilde{\mathbf{r}}$; 
$\mathbf{y}$ is the received signal vector; 
$\mathbf{s}$ is the transmit signal vector with the normalized power of unity and $P$ is the transmit power; 
$\mathbf{z} \sim \mathcal{CN}(0, \sigma^{2} \mathbf{I}_{N})$ is the additive white Gaussian noise (AWGN) at the Rx with average power $\sigma^{2}$.

In \eqref{equ:input-output}, the MIMO channel matrix can be explicitly expressed based on the field-response model \cite{Ma2024MIMO} as 
\begin{align}
\mathbf{H} \big( \tilde{\mathbf{r}}, \tilde{\mathbf{t}} \big) = {\mathbf{F}} \big( \tilde{\mathbf{r}} \big)^{H} {\mathbf{\Sigma}} {\mathbf{G}} \big( \tilde{\mathbf{t}} \big),
\end{align}
where ${\mathbf{G}} \big( \tilde{\mathbf{t}} \big) \triangleq \big[ {\mathbf{g}}( \mathbf{t}_{1}), {\mathbf{g}}( \mathbf{t}_{2}), \ldots, {\mathbf{g}}( \mathbf{t}_{M}) \big] \in \mathbb{C}^{L_{t} \times M}$ and 
${\mathbf{F}} \big( \tilde{\mathbf{r}} \big) \triangleq \big[ {\mathbf{f}}( \mathbf{r}_{1}), {\mathbf{f}}( \mathbf{r}_{2}), \ldots, {\mathbf{f}}( \mathbf{r}_{N}) \big] \in \mathbb{C}^{L_{r} \times N}$ are the field response matrices of $M$ T-MAs and $N$ R-MAs, respectively.  
$\bm{\Sigma} \in \mathbb{C}^{L_{r} \times L_{t}}$ denotes the path response matrix (PRM) between the Tx and Rx regions. 
$L_{r}$ and $L_{t}$ are the number of receive paths and transmit paths, respectively.
In ${\mathbf{G}} \big( \tilde{\mathbf{t}} \big)$, 
the field response vector of the $m$-th T-MA is given by
\begin{align}
	{\mathbf{g}}( \mathbf{t}_{m}) \triangleq 
	\big[ e^{\mathrm{j} \frac{2 \pi}{\lambda} \rho_{1}^{t}(\mathbf{t}_{m}) }, \ldots, e^{\mathrm{j} \frac{2 \pi}{\lambda} \rho_{L_{t}}^{t}(\mathbf{t}_{m}) } \big]^{T} \in \mathbb{C}^{L_{t} \times 1}, 
\end{align}
where $\lambda$ is the carrier wavelength;  $\rho_{l_{t}}^{t}(\mathbf{t}_{m}) = x_{m}^{t} \vartheta_{l_{t}}^{t} + y_{m}^{t} \varphi_{l_{t}}^{t}$ is the signal propagation distance difference between the $m$-th antenna position $\mathbf{t}_{m}$ and the coordinate origin for the $l_{t}$-th channel path;  
$\vartheta_{l_{t}}^{t} \triangleq  \cos\theta_{l_{t}}^{t} \sin\phi_{l_{t}}^{t}$, 
$\varphi_{l_{t}}^{t} \triangleq  \cos\phi_{l_{t}}^{t}$,
$\theta_{l_{t}}^{t}$ and $\phi_{l_{t}}^{t}$ are the azimuth and elevation AoDs, respectively.
Similarly, the field response vector of the $n$-th R-MA is given by
\begin{align}
	{\mathbf{f}}( \mathbf{r}_{n}) \triangleq 
	\big[ e^{\mathrm{j} \frac{2 \pi}{\lambda} \rho_{1}^{r}(\mathbf{r}_{n}) }, \ldots, e^{\mathrm{j} \frac{2 \pi}{\lambda} \rho_{L_{r}}^{r}(\mathbf{r}_{n}) } \big]^{T} \in \mathbb{C}^{L_{r} \times 1}, 
\end{align}
where $\rho_{l_{r}}^{r}(\mathbf{r}_{n}) = x_{n}^{r} \vartheta_{l_{r}}^{r} + y_{n}^{r} \varphi_{l_{r}}^{r}$ is the signal propagation distance difference between the antenna position $\mathbf{r}_{n}$ and the coordinate origin for the $l_{r}$-th channel path.  
$\vartheta_{l_{r}}^{r} \triangleq \cos\theta_{l_{r}}^{r} \sin\phi_{l_{r}}^{r}$, 
$\varphi_{l_{r}}^{r} \triangleq \cos\phi_{l_{r}}^{r}$,
$\theta_{l_{r}}^{r}$ and $\phi_{l_{r}}^{r}$ are the azimuth and elevation AoAs, respectively.

Our goal is to estimate the wireless channel at any pair of T-MA and R-MA positions in their corresponding movement regions $\mathcal{C}^{t}$ and $\mathcal{C}^{r}$. 
For the ease of antenna movement for channel estimation, we divide the two regions $\mathcal{C}^{t}$ and $\mathcal{C}^{r}$ into $G^{t} = A_{x} A_{y}/\Delta^{2}$ and $G^{r} = B_{x} B_{y}/\Delta^{2}$ grids, respectively, where $\Delta$ denotes the distance between the centers of adjacent grids at the Tx/Rx. Then, for simplicity, we assume that the T-/R-MAs only move over these grid centers while transmitting/receiving pilot signals for channel estimation. 
As at least one pilot symbol is needed for estimating the channels from a given Tx grid to $N$ Rx girds with $N$ R-MAs and there are in total $\binom{G^{t}}{M} \binom{G^{r}}{N}$ different pairs of Tx/Rx grids, at least $\binom{G^{t}}{M} \binom{G^{r}}{N} /N$ pilot symbols are needed to estimate all the channels over different Tx/Rx grid pairs.
This pilot overhead is practically prohibitive when the number of Tx and/or Rx grids is large for given $M$ and $N$. 
Fortunately, the field-response-based channels between all T-MA/R-MA position pairs share the same finite number of AoAs, AoDs, and associated complex gain coefficients, which yields intrinsic multidimensional signal characteristics, thus motivating us to leverage the tensor representation of received signals for efficient channel estimation. 
To achieve high channel reconstruction accuracy with low training overhead, we propose a tensor decomposition-based channel estimation/reconstruction method for MA-enabled MIMO systems in the next section.

\section{Tensor Decomposition-based Channel Estimation and Reconstruction}

In this section, we first introduce a two-stage Tx-Rx successive antenna movement pattern for pilot training such that the received pilot signals in both stages can be expressed as a third-order tensor. 
Then, we obtain the factor matrices of the tensor and estimate the parameters of multi-path channel components, including their AoDs, AoAs, and complex gain coefficients, thereby reconstructing the channel between arbitrary T-MA/R-MA positions. 
Finally, we analyze the uniqueness condition of the proposed tensor decomposition-based estimation/reconstruction method. 

\subsection{Tensor Representation}

In the first stage, we move the T-MAs over different positions to send pilot signals while maintaining the R-MAs fixed at an initial position denoted by $\tilde{\mathbf{r}}^{0} = \{\mathbf{r}_{1}^{0}, \ldots, \mathbf{r}_{N}^{0}\}$. 
Let $I = I_{x} I_{y}$ be the number of moved grid positions at T-MAs for pilot training, and the moved area $\mathcal{D}^{t}$ can be denoted as $I_{x} \Delta \times I_{y} \Delta$, where the associated moved positions are denoted by $\mathbf{t}_{i} = [x_{i}^{t}, y_{i}^{t}]^{T} \in \mathcal{D}^{t},  i=1, 
\ldots, I$.  
For each $\mathbf{t}_{i}$, one unit-power pilot symbol $s_{i}$ is transmitted by one of the T-MAs located at $\mathbf{t}_{i}$ (with the other T-MAs all switched off). The received pilot signals are thus given by
\begin{align}
	\mathbf{y}_{i}^{t} = \sqrt {P} {\mathbf{F}} \big( \tilde{\mathbf{r}}^{0} \big)^{H} {\mathbf{\Sigma}} {\mathbf{g}} \big( \mathbf{t}_{i} \big) s_{i} + \mathbf{z}_{i}^{t}, 
\end{align}
where $\mathbf{z}_{i}^{t}$ is the AWGN at the receiver. 
After multiplying the received signals with $s_{i}^{*}$ and concatenating them obtained with all $I$ different positions of T-MAs, we have 
\begin{align}
\mathbf{Y}^{t} & = \sqrt {P} {\mathbf{F}} \big( \tilde{\mathbf{r}}^{0} \big)^{H} {\mathbf{\Sigma}} {\mathbf{G}} \big( \mathcal{D}^{t} \big) + \mathbf{Z}^{t}, \label{equ:y_t}
\end{align}
where $\mathbf{Y}^{t} = [\mathbf{y}_{1}^{t}, \ldots, \mathbf{y}_{I}^{t}] \in \mathbb{C}^{N \times I}$, 
$\mathbf{Z}^{t} = [\mathbf{z}_{1}^{t}, \ldots, 	\mathbf{z}_{I}^{t}] \in \mathbb{C}^{N \times I}$, 
and ${\mathbf{G}} ( \mathcal{D}^{t} ) \triangleq [{\mathbf{g}} ( \mathbf{t}_{1} ), \ldots, {\mathbf{g}} ( \mathbf{t}_{I} )]   \in \mathbb{C}^{L_{t} \times I}$. 
Note that each row of ${\mathbf{G}}$ corresponds to the AoDs from different positions of T-MAs.
We define $\mathbf{A}(\bm{\vartheta}^{t}, \bm{\varphi}^{t}) = {\mathbf{G}} ( \mathcal{D}^{t} )^{H}$, 
where $\bm{\vartheta}^{t} = [\vartheta_{1}^{t}, \ldots, \vartheta_{L_{t}}^{t}]$, 
$\bm{\varphi}^{t} = [\varphi_{1}^{t}, \ldots, \varphi_{L_{t}}^{t}]$, and the $l_{t}$-th column of $\mathbf{A}(\bm{\vartheta}^{t}, \bm{\varphi}^{t})$ can be regarded as a Tx-side steering vector given by
\begin{align}
\mathbf{a}(\vartheta_{l_{t}}^{t}, \varphi_{l_{t}}^{t}) & = 
\big[e^{-\mathrm{j} \frac{2 \pi}{\lambda} (x_{1}^{t} \vartheta_{l_{t}}^{t} + y_{1}^{t} \varphi_{l_{t}}^{t}) }, \ldots, e^{-\mathrm{j} \frac{2 \pi}{\lambda} (x_{I}^{t} \vartheta_{l_{t}}^{t} + y_{I}^{t} \varphi_{l_{t}}^{t}) } \big]^{T} \nonumber \\
& = \mathbf{a}_{x}(\vartheta_{l_{t}}^{t}) \otimes 
\mathbf{a}_{y}(\varphi_{l_{t}}^{t}),
\end{align}
where $\mathbf{a}_{x}(\vartheta_{l_{t}}^{t}) \in \mathbb{C}^{I_{x} \times 1}$ and
$\mathbf{a}_{y}(\varphi_{l_{t}}^{t}) \in \mathbb{C}^{I_{y} \times 1}$. 
Then, the received signal matrix can be rewritten as
\begin{align}
\overline{\mathbf{Y}}^{t} 
& = \sqrt {P} \mathbf{A}({\bm{\vartheta}^{t}, \bm{\varphi}^{t}}) {\mathbf{\Sigma}}^{H} {\mathbf{F}} \big( \tilde{\mathbf{r}}^{0} \big)   + \overline{\mathbf{Z}}^{t}  \nonumber \\ 
& = (\mathbf{A}_{x} \odot \mathbf{A}_{y}) \mathbf{D}_{t}^{T}  + \overline{\mathbf{Z}}^{t}, \label{equ:y_t_line}
\end{align}
where 
$\overline{\mathbf{Y}}^{t} = (\mathbf{Y}^{t})^{H}  \in \mathbb{C}^{I \times N}$, 
$\overline{\mathbf{Z}}^{t} = (\mathbf{Z}^{t})^{H}$,
$\mathbf{D}_{t} = \big( \sqrt {P} {\mathbf{\Sigma}}^{H} {\mathbf{F}} ( \tilde{\mathbf{r}}^{0} ) \big)^{T}$, 
$\mathbf{A}_{x} = [\mathbf{a}_{x}(\vartheta_{1}^{t}), \ldots, \mathbf{a}_{x}(\vartheta_{L_{t}}^{t})]$ and 
$\mathbf{A}_{y} = [\mathbf{a}_{y}(\varphi_{1}^{t}), \ldots, \mathbf{a}_{y}(\varphi_{L_{t}}^{t})]$.

In the second stage, we keep the T-MAs fixed at the initial position denoted by 
$\tilde{\mathbf{t}}^{0} = \{\mathbf{t}_{1}^{0}, \ldots, \mathbf{t}_{M}^{0}\}$ while moving R-MAs over different positions to receive pilot signals. 
Let $\mathcal{D}^{r}$ be the moved area for R-MAs denoted by $J_{x} \Delta \times J_{y} \Delta$, where $J = J_{x} J_{y}$ is the total number of moved grid positions at R-MAs for pilot training and the associated moved positions of $N$ R-MAs are denoted by $\tilde{\mathbf{r}}_{j} = [\mathbf{r}_{1,j}, \ldots, \mathbf{r}_{N,j}] $, 
$\mathbf{r}_{n,j} = [x_{n,j}^{r}, y_{n,j}^{r}]^{T} \in \mathcal{D}^{r},  j=1, 
\ldots, J/N$. 
Let $\mathbf{S}$ be the pilot matrix transmitted by all T-MAs at the same time and satisfy $\mathbf{S} \mathbf{S}^{H} = \mathbf{I}_{M}$. Then, the received signals are given by
\begin{align}
	\mathbf{Y}_{j}^{r} = \sqrt {P} {\mathbf{F}} \big( \tilde{\mathbf{r}}_{j} \big)^{H} {\mathbf{\Sigma}} {\mathbf{G}} \big( \tilde{\mathbf{t}}^{0} \big) \mathbf{S} + \mathbf{Z}_{j}^{r}, 
\end{align}
where $\mathbf{Z}_{j}^{r}$ is the AWGN matrix. 
In this stage, we move R-MAs $J/N$ times to cover the area $\mathcal{D}^{r}$. 
After performing $\overline{\mathbf{Y}}_{j}^{r} = \mathbf{Y}_{j}^{r} \mathbf{S}^{H}$ and concatenating $\overline{\mathbf{Y}}_{j}^{r}$ of $J/N$ moves of R-MAs, we have 
\begin{align}
	\overline{\mathbf{Y}}^{r} 
	& = \sqrt {P} {\mathbf{F}} \big( \mathcal{D}^{r} \big)^{H} {\mathbf{\Sigma}} {\mathbf{G}} \big( \tilde{\mathbf{t}}^{0} \big)  + \overline{\mathbf{Z}}^{r} \nonumber \\
	& = (\mathbf{B}_{x} \odot \mathbf{B}_{y}) \mathbf{D}_{r}^{T}  + \overline{\mathbf{Z}}^{r}, \label{equ:y_r_line}
\end{align}
where $\overline{\mathbf{Y}}^{r} = \big[ (\overline{\mathbf{Y}}_{1}^{r})^{T}, \ldots, (\overline{\mathbf{Y}}_{J/N}^{r})^{T} \big]^{T} \in \mathbb{C}^{J \times M}$, 
$\overline{\mathbf{Z}}^{r} = \big[ (\mathbf{Z}_{1}^{r})^{T}, \ldots, (\mathbf{Z}_{J/N}^{r})^{T} \big]^{T}$, 
the second equality follows from defining $\mathbf{D}_{r} = \big( \sqrt {P} {\mathbf{\Sigma}} {\mathbf{G}} ( \tilde{\mathbf{t}}^{0} ) \big)^{T}$ and 
${\mathbf{F}} ( \mathcal{D}^{r} )^{H} \triangleq \mathbf{B}(\bm{\vartheta}^{r}, \bm{\varphi}^{r}) = \mathbf{B}_{x} \odot \mathbf{B}_{y}$, with
$\bm{\vartheta}^{r} = [\vartheta_{1}^{r}, \ldots, \vartheta_{L_{r}}^{r}]$ and 
$\bm{\varphi}^{r} = [\varphi_{1}^{r}, \ldots, \varphi_{L_{r}}^{r}]$.

Based on the definition of CP decomposition \cite{kolda2009tensor}, both $\overline{\mathbf{Y}}^{t}$ in \eqref{equ:y_t_line} and $\overline{\mathbf{Y}}^{r}$ in \eqref{equ:y_r_line} can be regarded as a matrix form of a third-order tensor $\bm{\overline{\mathcal{Y}}}^{t} \in \mathbb{C}^{I_{y} \times I_{x} \times N}$ or 
$\bm{\overline{\mathcal{Y}}}^{r} \in \mathbb{C}^{J_{y} \times J_{x} \times M}$. 
For $\bm{\overline{\mathcal{Y}}}^{t}$, its
$(i_{y}, i_{x}, n)$-th entry is the $n$-th row and the $((i_{x}-1)I_{y} + i_{y})$-th column entry of $\overline{\mathbf{Y}}^{t}$, and the three factor matrices are $\lbrace \mathbf{A}_{y}$, $\mathbf{A}_{x}$, $\mathbf{D}_{t} \rbrace$.  
Similarly, for $\bm{\overline{\mathcal{Y}}}^{r}$, its three factor matrices are $\lbrace \mathbf{B}_{y}$, $\mathbf{B}_{x}$, $\mathbf{D}_{r} \rbrace$.  
From \eqref{equ:y_t_line} and \eqref{equ:y_r_line}, we can observe that the factor matrices of the tensors $\bm{\overline{\mathcal{Y}}}^{t}$ and $\bm{\overline{\mathcal{Y}}}^{r}$ incorporate all the channel information, which motivates us to leverage the tensor decomposition for channel estimation/reconstruction in the next.

\subsection{Channel Estimation and Reconstruction} 

\subsubsection{Estimation of Factor Matrices}
Given the received signal tensor $\bm{\overline{\mathcal{Y}}}^{t}$ or $\bm{\overline{\mathcal{Y}}}^{r}$, there are various methods to solve the CP decomposition problem for estimating the factor matrices. 
One well-known method is the alternating least squares (ALS), where one factor matrix is updated by fixing the other factor matrices alternately in an iterative manner until the convergence is reached \cite{Zhang2022Tensor}. 
For instance, for the CP decomposition of $\bm{\overline{\mathcal{Y}}}^{t}$, we have
\begin{align}
	\mathbf{A}_{y}^{(k+1)} & \!= \arg\min_{\mathbf{A}_{y}} \Big\lVert \overline{\mathbf{Y}}_{(1)}^{t} \!-\! \mathbf{A}_{y} ( \mathbf{D}_{t}^{(k)} \odot \mathbf{A}_{x}^{(k)} )^{T} \Big\rVert_{F}^{2},  \label{equ:ALS1} \\
	\mathbf{A}_{x}^{(k+1)} & \!= \arg\min_{\mathbf{A}_{x}} \Big\lVert \overline{\mathbf{Y}}_{(2)}^{t} \!-\! \mathbf{A}_{x} ( \mathbf{D}_{t}^{(k)} \odot \mathbf{A}_{y}^{(k+1)} )^{T} \Big\rVert_{F}^{2}, \label{equ:ALS2} \\
	\mathbf{D}_{t}^{(k+1)} & \!= \arg\min_{\mathbf{D}_{t}} \Big\lVert \overline{\mathbf{Y}}_{(3)}^{t} \!-\! \mathbf{D}_{t} ( \mathbf{A}_{x}^{(k+1)} \odot \mathbf{A}_{y}^{(k+1)} )^{T} \Big\rVert_{F}^{2}, \label{equ:ALS3}
\end{align}
where $\overline{\mathbf{Y}}_{(1)}^{t}$, $\overline{\mathbf{Y}}_{(2)}^{t}$, and $\overline{\mathbf{Y}}_{(3)}^{t}$ are the mode-1, mode-2, and mode-3 unfolding of $\bm{\overline{\mathcal{Y}}}^{t}$, 
while $\mathbf{A}_{y}^{(k)}$, $\mathbf{A}_{x}^{(k)}$, and $\mathbf{D}_{t}^{(k)}$ are the updated factor matrices at the $k$-th iteration.  
By conducting the above three updates alternately in an iterative manner, we can obtain the estimation of factor matrices as $\lbrace \hat{\mathbf{A}}_{y}$, $\hat{\mathbf{A}}_{x}$, $\hat{\mathbf{D}}_{t} \rbrace$ after convergence. 
In a similar way as \eqref{equ:ALS1}-\eqref{equ:ALS3}, we can obtain the estimated factor matrices of $\bm{\overline{\mathcal{Y}}}^{r}$, which are denoted as $\lbrace \hat{\mathbf{B}}_{y}$, $\hat{\mathbf{B}}_{x}$, $\hat{\mathbf{D}}_{r} \rbrace$.

\subsubsection{Estimation of AoDs and AoAs}
We use the estimated factor matrices to estimate AoDs and AoAs.
Note that via CP decomposition, the estimated factor matrices are generally not unbiased and subjected to two types of ambiguities, i.e., scaling ambiguity and permutation ambiguity \cite{kolda2009tensor}. 
For the estimated factor matrices $\lbrace \hat{\mathbf{A}}_{y}$, $\hat{\mathbf{A}}_{x}$, $\hat{\mathbf{D}}_{t} \rbrace$, they satisfy the following relationship
\begin{align}
	\hat{\mathbf{A}}_{y} & = \mathbf{A}_{y} \mathbf{\Pi}^{t} \mathbf{\Lambda}_{y}^{t} + \mathbf{E}_{y}^{t}, \label{equ:Ay}\\ 
	\hat{\mathbf{A}}_{x} & = \mathbf{A}_{x} \mathbf{\Pi}^{t} \mathbf{\Lambda}_{x}^{t} + \mathbf{E}_{x}^{t}, \label{equ:Ax}\\ 
	\hat{\mathbf{D}}_{t} &= \mathbf{D}_{t} \mathbf{\Pi}^{t} \mathbf{\Lambda}_{t}^{t} + \mathbf{E}_{t}^{t}, \label{equ:Dt}
\end{align} 
where 
$\mathbf{\Pi}^{t}$ denotes a permutation matrix, 
$\mathbf{\Lambda}_{y}^{t}$, $\mathbf{\Lambda}_{x}^{t}$, and $\mathbf{\Lambda}_{t}^{t}$ are the diagonal matrices,
$\mathbf{E}_{y}^{t}$, $\mathbf{E}_{x}^{t}$, and $\mathbf{E}_{t}^{t}$ denote the errors of tensor decomposition. 
From \eqref{equ:Ay} to \eqref{equ:Dt}, we can observe that the estimates of $\lbrace \hat{\mathbf{A}}_{y}$, $\hat{\mathbf{A}}_{x}$, $\hat{\mathbf{D}}_{t} \rbrace$ are not exactly the same as the original factor matrices, even in the noiseless case. 
Fortunately, the ambiguities do not disrupt the pairing relation among the estimated and true factor matrices, allowing the AoDs to be extracted from the corresponding column vector of factor matrices. 
Moreover, benefitting from the successive antenna movement pattern, the factor matrices $\mathbf{A}_{y}$ and $\mathbf{A}_{x}$ exhibit the inherent Vandermonde structure, which facilitates estimating AoDs in a gridless manner.
We take $\mathbf{A}_{x}$ as an example to illustrate the process of AoD estimation. 
From \eqref{equ:Ax}, we have
\begin{align}
\hat{\mathbf{a}}_{x,l_{t}} & = \delta_{x,l_{t}} \mathbf{a}_{x}(\vartheta_{l_{t}}^{t}) + \mathbf{e}_{x,l_{t}}, \label{equ:ax}
\end{align}
where $\delta_{x,l_{t}}^{t}$ is the $l_{t}$-th diagonal entry of $\mathbf{\Lambda}_{x}^{t}$, 
$\hat{\mathbf{a}}_{x,l_{t}}$ and $\mathbf{e}_{x,l_{t}}^{t}$ are the $l_{t}$-th column of $\hat{\mathbf{A}}_{x}$ and $\mathbf{E}_{x}^{t}$, respectively. 
Note that $\mathbf{a}_{x}(\vartheta_{l_{t}}^{t})$ is the $l_{t}$-th column of $\mathbf{A}_{x}$ expressed as
$\mathbf{a}_{x}(\vartheta_{l_{t}}^{t}) = \big[1, \omega_{l_{t}}^{t}, \ldots, (\omega_{l_{t}}^{t})^{I_{x}-1} \big]^{T}$ with the generator
$\omega_{l_{t}}^{t} = e^{-\mathrm{j} \frac{2 \pi \Delta}{\lambda}  \vartheta_{l_{t}}^{t}}$.
In \eqref{equ:ax}, due to the Vandermonde structure of $\mathbf{A}_{x}$, we can further obtain $[\hat{\mathbf{a}}_{x,l_{t}}]_{2:I_{x}} = \omega_{l_{t}}^{t} [\hat{\mathbf{a}}_{x,l_{t}}]_{1:I_{x}-1}$ in the noiseless case.  
Due to the one-to-one mapping between $\vartheta_{l_{t}}^{t}$ and $\omega_{l_{t}}^{t}$, 
the estimation of $\vartheta_{l_{t}}^{t}$ can be obtained from the phase of $\omega_{l_{t}}^{t}$ as
\begin{align}
\hat{\vartheta}_{l_{t}}^{t} = - \frac{\lambda}{2\pi \Delta}  \angle  \hat{\omega}_{l_{t}}^{t}, \ l_{t} = 1, \ldots, L_{t},  \label{equ:theta_t_est}
\end{align}
where $\angle$ denotes the operator of phase angle extraction, 
$\hat{\omega}_{l_{t}}^{t} =  [\hat{\mathbf{a}}_{x,l_{t}}]_{1:I_{x}-1}^{\dagger} [\hat{\mathbf{a}}_{x,l_{t}}]_{2:I_{x}} $ is the estimated generator.
Similarly, we can use the factor matrix $\hat{\mathbf{A}}_{y}$ to obtain the estimate of $\varphi_{l_{t}}^{t}$ as
\begin{align}
	\hat{\varphi}_{l_{t}}^{t} = - \frac{\lambda}{2\pi \Delta}  \angle  \hat{\upsilon}_{l_{t}}^{t}, \ l_{t} = 1, \ldots, L_{t},  \label{equ:phi_t_est}
\end{align}
where $\hat{\upsilon}_{l_{t}}^{t} =  [\hat{\mathbf{a}}_{y,l_{t}}]_{1:I_{y}-1}^{\dagger} [\hat{\mathbf{a}}_{y,l_{t}}]_{2:I_{y}} $, and  $\hat{\mathbf{a}}_{y,l_{t}}$ is the $l_{t}$-th column of $\hat{\mathbf{A}}_{y}$.

In terms of AoAs estimation, 
we can employ the same method as that for the AoDs estimation above. 
To be specific, the estimated AoAs  
$\{\hat{\vartheta}_{l_{r}}^{r}\}_{l_{r}=1}^{L_{r}}$ and
$\{\hat{\varphi}_{l_{r}}^{r}\}_{l_{r}=1}^{L_{r}}$ can be respectively given by
\begin{align}
	\hat{\vartheta}_{l_{r}}^{r} = - \frac{\lambda}{2\pi \Delta}  \angle  \hat{\omega}_{l_{r}}^{r}, \ l_{r} = 1, \ldots, L_{r}, \label{equ:theta_r_est}
\end{align}
\begin{align}
	\hat{\varphi}_{l_{r}}^{r} = - \frac{\lambda}{2\pi \Delta}  \angle  \hat{\upsilon}_{l_{r}}^{r}, \ l_{r} = 1, \ldots, L_{r}, \label{equ:phi_r_est}
\end{align}
where $\hat{\omega}_{l_{r}}^{r} = [\hat{\mathbf{b}}_{x,l_{r}}]_{1:J_{x}-1}^{\dagger} [\hat{\mathbf{b}}_{x,l_{r}}]_{2:J_{x}} $ and 
$\hat{\upsilon}_{l_{r}}^{r} = 
[\hat{\mathbf{b}}_{y,l_{r}}]_{1:J_{y}-1}^{\dagger} [\hat{\mathbf{b}}_{y,l_{r}}]_{2:J_{y}} $, 
$\hat{\mathbf{b}}_{x,l_{r}}$ and $\hat{\mathbf{b}}_{y,l_{r}}$ are the $l_{r}$-th column of $\hat{\mathbf{B}}_{x}$ and $\hat{\mathbf{B}}_{y}$, respectively.

\begin{algorithm}[!tp]
	\renewcommand{\algorithmicrequire}{\textbf{Input:}}  
	\renewcommand{\algorithmicensure}{\textbf{Output:}}  
	\caption{Proposed Tensor Decomposition-Based Channel Estimation and Reconstruction}
	\label{alg:tensor decomposition}
	\begin{algorithmic}[1]
		\Require
		Received signals $\bm{\overline{\mathcal{Y}}}^{t}$ and $\bm{\overline{\mathcal{Y}}}^{r}$,
		moved area $\mathcal{D}^{t}$ and $\mathcal{D}^{r}$.
		
		\State Estimate factor matrices $\lbrace \hat{\mathbf{A}}_{y}$, $\hat{\mathbf{A}}_{x}$, $\hat{\mathbf{D}}_{t} \rbrace$ and $\lbrace \hat{\mathbf{B}}_{y}$, $\hat{\mathbf{B}}_{x}$, $\hat{\mathbf{D}}_{r} \rbrace$ based on CP decomposition;  
		
		\State Estimate AoDs $\lbrace \hat{\vartheta}_{l_{t}}^{t}, \hat{\varphi}_{l_{t}}^{t} \rbrace_{l_{t}=1}^{L_{t}}$ via \eqref{equ:theta_t_est} and \eqref{equ:phi_t_est}; \
		\State Estimate AoAs $\lbrace \hat{\vartheta}_{l_{r}}^{r}, \hat{\varphi}_{l_{r}}^{r} \rbrace_{l_{r}=1}^{L_{r}}$ via  \eqref{equ:theta_r_est} and \eqref{equ:phi_r_est}; \
		\State Estimate $\hat{\bm{\gamma}}$ via \eqref{equ:gamma}; \ 
		\State Reconstruct channel matrix $\hat{\mathbf{H}} ( \tilde{\mathbf{r}}, \tilde{\mathbf{t}} )$ via \eqref{equ:ChannelReconstrct}; \
		
		\Ensure
		$\hat{\mathbf{H}} ( \tilde{\mathbf{r}}, \tilde{\mathbf{t}} )$.
	\end{algorithmic}
\end{algorithm}

\subsubsection{Channel Reconstruction}
With the estimated AoDs and AoAs, we now turn to the estimation of PRM $\mathbf{\Sigma}$ and then reconstruct the channel between any pair of T-MA and R-MA positions.
From \eqref{equ:y_t} and \eqref{equ:y_r_line}, considering the vectorization of the received signals $\mathbf{Y}^{t}$ and $\overline{\mathbf{Y}}^{r}$, we have 
\begin{align}
\begin{cases}
	\mathbf{y}^{t} = \bm{\Phi}^{t}\big(\bm{\vartheta}^{t}, \bm{\varphi}^{t}, \bm{\vartheta}^{r}, \bm{\varphi}^{r} \big) \bm{\gamma}  + \mathbf{z}^{t}, \\
	\overline{\mathbf{y}}^{r} = \bm{\Phi}^{r}\big(\bm{\vartheta}^{t}, \bm{\varphi}^{t}, \bm{\vartheta}^{r}, \bm{\varphi}^{r} \big) \bm{\gamma}  + \overline{\mathbf{z}}^{r},
\end{cases}
\end{align}
where $\mathbf{y}^{t} = \text{vec}\big(\mathbf{Y}^{t}\big) \in \mathbb{C}^{IN \times 1}$ and 
$\overline{\mathbf{y}}^{r} = \text{vec}\big(\overline{\mathbf{Y}}^{r} \big) \in \mathbb{C}^{JM \times 1}$,  $\bm{\gamma} = \text{vec}\big({\mathbf{\Sigma}}\big) \in \mathbb{C}^{L_{r}L_{t} \times 1}$ denotes the vectorization of $\mathbf{\Sigma}$,  
$\mathbf{z}^{t} = \text{vec}\big(\mathbf{Z}^{t}\big)$ and 
$\overline{\mathbf{z}}^{r} = \text{vec}\big(\overline{\mathbf{Z}}^{r} \big)$ are the associated noise vectors,
$\bm{\Phi}^{t}\big(\bm{\vartheta}^{t}, \bm{\varphi}^{t}, \bm{\vartheta}^{r}, \bm{\varphi}^{r} \big) = \sqrt {P}  \big( {\mathbf{G}} ( \mathcal{D}^{t} )^{T}  \otimes {\mathbf{F}} ( \tilde{\mathbf{r}}^{0} )^{H} \big)$, and
$\bm{\Phi}^{r}\big(\bm{\vartheta}^{t}, \bm{\varphi}^{t}, \bm{\vartheta}^{r}, \bm{\varphi}^{r} \big) = \sqrt {P}  \big( {\mathbf{G}} ( \tilde{\mathbf{t}}^{0} )^{T}  \otimes {\mathbf{F}} ( \mathcal{D}^{r} )^{H} \big)$. 
Then, we can obtain the estimation of $\bm{\gamma}$ as
\begin{align}
\hat{\bm{\gamma}} = \hat{\bm{\Phi}}^{\dagger} \mathbf{y}, \label{equ:gamma}
\end{align}
where $\mathbf{y}= [(\mathbf{y}^{t})^{T}, (\overline{\mathbf{y}}^{r})^{T} ]^{T}$, 
$\hat{\bm{\Phi}} = \big[(\hat{\bm{\Phi}}^{t})^{T}, (\hat{\bm{\Phi}}^{r})^{T}\big]^{T}$, 
$\hat{\bm{\Phi}}^{t}$ and $\hat{\bm{\Phi}}^{r}$ are constructed based on the estimated AoDs and AoAs $\lbrace \hat{\vartheta}_{l_{t}}^{t}, \hat{\varphi}_{l_{t}}^{t} \rbrace_{l_{t}=1}^{L_{t}}$ and
$\lbrace \hat{\vartheta}_{l_{r}}^{r}, \hat{\varphi}_{l_{r}}^{r} \rbrace_{l_{r}=1}^{L_{r}}$.
Finally, with the estimated channel parameters, the channel for any set of T-MA positions $\tilde{\mathbf{t}}$ and R-MA positions $\tilde{\mathbf{r}}$ can be reconstructed as
\begin{align}
\hat{\mathbf{H}} \big( \tilde{\mathbf{r}}, \tilde{\mathbf{t}} \big) = \hat{\mathbf{F}} \big( \tilde{\mathbf{r}} \big)^{H} \hat{\mathbf{\Sigma}} \hat{\mathbf{G}} \big( \tilde{\mathbf{t}} \big),  
\label{equ:ChannelReconstrct}
\end{align}
where $\hat{\mathbf{\Sigma}}$ is the matrix version of  of $\hat{\bm{\gamma}}$, 
$\hat{\mathbf{F}} \big( \tilde{\mathbf{r}} \big)$ and 
$\hat{\mathbf{G}} \big( \tilde{\mathbf{t}} \big)$ are the estimated field response matrices based on the estimated AoAs and AoDs, respectively. 
The overall algorithm above is summarized in Algorithm \ref{alg:tensor decomposition}.
 
\subsection{Uniqueness Condition of Tensor Decomposition}
\label{subsec:uniqueness}
The uniqueness of the tensor decomposition is essential for the considered channel estimation problem, ensuring that the decomposed factor matrices can be used to recover all unknown multi-path channel parameters. 
We first introduce the general sufficient condition for uniqueness of CP decomposition \cite{Stegeman2007On}, which is given as follows. 

\begin{lemma} \label{lemma:Kruskal}
	Let $\mathbf{A}^{(1)} \in \mathbb{C}^{I_{1} \times L}$, $\mathbf{A}^{(2)}\in \mathbb{C}^{I_{2} \times L}$, and $\mathbf{A}^{(3)}\in \mathbb{C}^{I_{3} \times L}$ be the factor matrices of a third-order tensor $\bm{\mathcal{X}} \in \mathbb{C}^{I_{1} \times I_{2} \times I_{3}}$ with rank $L$. 
	If the following inequality holds,
	\begin{equation}
	\min(I_{1}, L) + \min(I_{2}, L) + \min(I_{3}, L)  \geq 2L + 2, \label{equ:uniqueness}
    \end{equation}
	then the CP decomposition of $\bm{\mathcal{X}}$ is unique. 
\end{lemma}

Lemma \ref{lemma:Kruskal} provides the condition of factor matrices guaranteeing the uniqueness of CP decomposition, which can be used to analyze the considered channel estimation problem with tensors $\bm{\overline{\mathcal{Y}}}^{t}$ and $\bm{\overline{\mathcal{Y}}}^{r}$.
For instance, to acquire a unique decompostion of  $\bm{\overline{\mathcal{Y}}}^{t}$, the dimension of its factor matrices $\lbrace \mathbf{A}_{y}$, $\mathbf{A}_{x}$, $\mathbf{D}_{t} \rbrace$ should satisfy 
$\min(I_{x}, L_{t}) + \min(I_{y}, L_{t}) + \min(N, L_{t}) \geq 2 L_{t} + 2$. 
It can be observed that for a given number of channel paths $L_{t}$, the required numbers of T-MA movement $I_{x}$ and $I_{y}$ are only related to $L_{t}$, but independent of the total number of grid positions in the Tx-side antenna movement region. 
Similar result can also be obtained for R-MAs.  
Thus, the proposed channel estimation method can acquire the information required for complete channel reconstruction by moving T-MAs/R-MAs over a finite number of positions for channel measurement, thus significantly reducing the pilot training overhead.

\section{Simulation Results}
\label{Sec:Simulation}
In this section, we present simulation results to evaluate the proposed tensor decomposition-based estimation/reconstruction method for MA-enabled MIMO systems. 
In simulation, the number of T-MAs/R-MAs is set to $M=N=4$ and the local regions $\mathcal{C}^{t}$ and $\mathcal{C}^{r}$ are both of size $8 \lambda \times 8 \lambda$. 
The numbers of Tx and Rx channel paths are set to $L_{t} = L_{r} = 3$, 
with each diagonal element of PRM ${\mathbf{\Sigma}}$ generated from $\mathcal{CN}(0,\eta/((\eta+1)L_{r}))$
and each non-diagonal element of ${\mathbf{\Sigma}}$ generated from 
$\mathcal{CN}(0,1/((\eta+1)(L_{r}-1)L_{r}))$, 
where $\eta$ denotes the average power ratio of diagonal elements to non-diagonal elements and is set to $1$ \cite{Ma2023Compressed}. 
The azimuth and elevation AoDs/AoAs are uniformly generated from the distribution $[0, \pi]$. 
As we uniformly divide both $\mathcal{C}^{t}$ and $\mathcal{C}^{r}$ into $G^{t}$ and $G^{r}$ grids, respectively, we set the distance of adjacent grids equal to $\Delta = \lambda/5$. 
Then, the true and reconstructed channels between all different grids in $\mathcal{C}^{t}$ and $\mathcal{C}^{r}$ can be expressed as the matrices $\mathbf{H} \in \mathbb{C}^{G^{r} \times G^{t}}$ and $\hat{\mathbf{H}} \in \mathbb{C}^{G^{r} \times G^{t}}$, respectively. 
Moreover, we define $\beta^{t} \triangleq I/G^{t}$ and $\beta^{r} \triangleq J/G^{r}$ as the portion of the area in $\mathcal{D}^{t}$ and $\mathcal{D}^{r}$ that is visited by T-MAs/R-MAs for pilot training, respectively.
In addition, we set $I_{x} = I_{y}$ and $J_{x} = J_{y}$ in simulation.   
The initial positions $\tilde{\mathbf{t}}^{0}$ and $\tilde{\mathbf{r}}^{0}$ for T-MAs and R-MAs are set to the vertexes of $\mathcal{C}^{t}$ and $\mathcal{C}^{r}$, respectively. 
The normalized mean square error (NMSE) for channel reconstruction is given by $\text{NMSE} = \big\lVert \mathbf{H} - \hat{\mathbf{H}} \big\rVert_{F}^{2}/ \big\lVert \mathbf{H} \big\rVert_{F}^{2}$. 
The signal-to-noise ratio (SNR) is defined as $P / \sigma^{2}$. 
Moreover, the Cramér-Rao bound (CRB) \cite{kay1993fundamentals} and the successive transmitter-receiver CS scheme in \cite{Ma2023Compressed} are considered as benchmarks, for which the orthogonal matching pursuit (OMP) and simultaneous OMP (SOMP) algorithms \cite{Zhang2022MMV} are adopted.

\begin{figure}[!tp]
	\centering
	\includegraphics[width=2.5in]
	{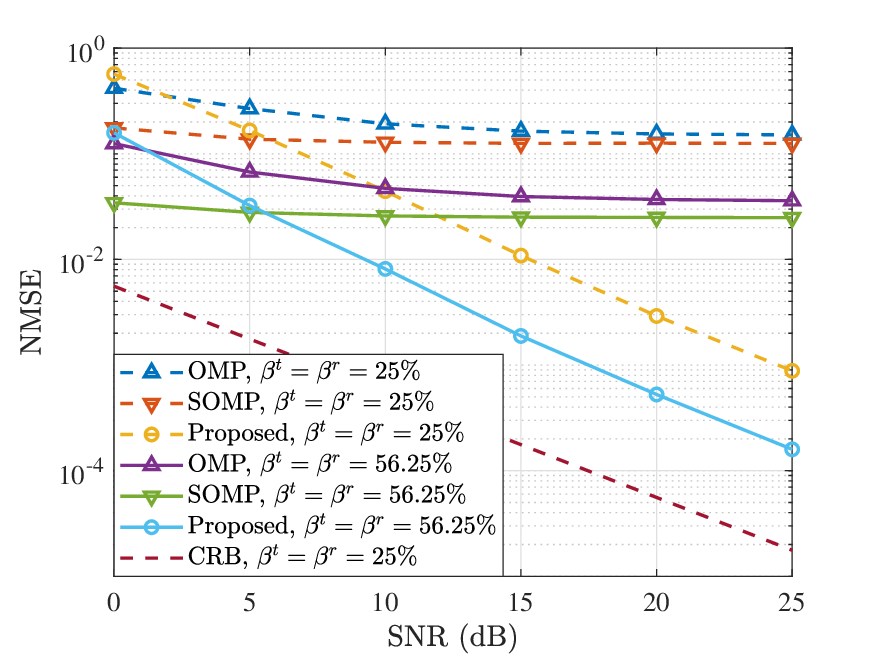}
	\caption{NMSE of channel reconstruction versus SNR.}
	\label{Fig_VaryingSNR_NMSE}
\end{figure}

\begin{figure}[!tp]
	\centering
	\includegraphics[width=2.5in]
	{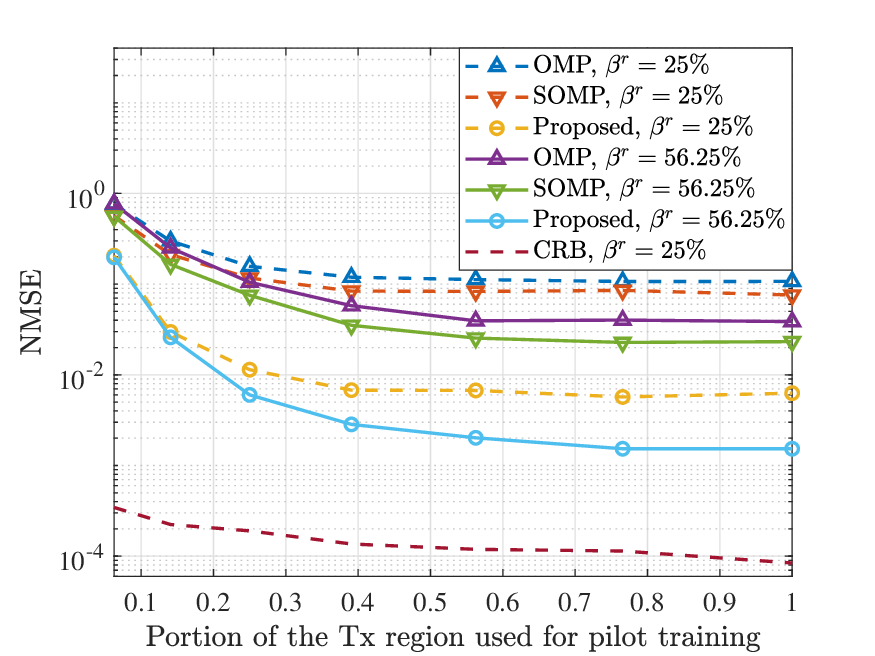}
	\caption{NMSE of channel reconstruction versus $\beta^{t}$.}
	\label{Fig_VaryingT_NMSE}
\end{figure}

Fig. \ref{Fig_VaryingSNR_NMSE} shows the NMSE of reconstructed channels versus SNR, with $\beta^{t} = \beta^{r} = 25 \%$ or $56.25 \%$.   
It is observed that the proposed method achieves lower NMSE than the two CS-based algorithms.
With the increase of SNR, the NMSE of the proposed method continues to decrease, while those of the benchmark algorithms get saturated. 
This is due to the uniqueness of tensor decomposition and the fact that we are able to directly estimate channel parameters without any discretization.
Moreover, the NMSE of the proposed method and CRB exhibit a similar trend, while their gap can be reduced by increasing the pilot training overhead.

Fig. \ref{Fig_VaryingT_NMSE} depicts the NMSE versus the portion of the Tx region used for pilot training, $\beta^{t}$, where SNR is set to $15$ dB, and $\beta^{r}$ is set to $25 \%$ or $56.25 \%$.
As the moved area for pilot training increases, the channel estimation performance of the proposed method improves and is much better compared to the benchmark algorithms.
For example, compared to the benchmark algorithms with $\beta^{t} = 100\%$, the proposed method can achieve comparable NMSE with $\beta^{t} = 20\%$, thus achieving $80\%$ reduction in pilot overhead.

\section{Conclusion}

In this letter, we proposed a new tensor decomposition-based channel estimation/reconstruction method for MA-enabled MIMO systems. 
We first introduced a two-stage Tx-Rx successive antenna movement pattern for pilot training. 
Then, we obtained the factor matrices of the tensor and estimated the parameters of multi-path channel components, thereby reconstructing the complete channel between arbitrary Tx/Rx MA positions.
The tensor decomposition uniqueness condition analysis and simulation results demonstrated that the proposed method can achieve significant pilot overhead reduction with the same channel estimation accuracy as compared to benchmark schemes.




\begin{thebibliography}{10}
	
\bibitem{lu2014overview}
L.~Lu, G.~Y. Li, A.~L. Swindlehurst, A.~Ashikhmin, and R.~Zhang, ``{An overview
	of massive MIMO: Benefits and challenges},'' \emph{IEEE J. Sel. Topics Signal
	Process.}, vol.~8, no.~5, pp. 742--758, Oct. 2014.

\bibitem{Zhang2024TWC}
R.~Zhang, L.~Cheng, S.~Wang, Y.~Lou, Y.~Gao, W.~Wu, and D.~W.~K. Ng,
``Integrated sensing and communication with massive {MIMO}: A unified tensor
approach for channel and target parameter estimation,'' \emph{IEEE Trans.
	Wireless Commun.}, vol.~23, no.~8, pp. 8571--8587, Aug. 2024.

\bibitem{Zhu2023Movable_Mag}
L.~Zhu, W.~Ma, and R.~Zhang, ``Movable antennas for wireless communication:
Opportunities and challenges,'' \emph{IEEE Commun. Mag.}, vol.~62, no.~6, pp.
114--120, Jun. 2024.

\bibitem{Wong2021Fluid}
K.-K. Wong, A.~Shojaeifard, K.-F. Tong, and Y.~Zhang, ``Fluid antenna
systems,'' \emph{IEEE Trans. Wireless Commun.}, vol.~20, no.~3, pp.
1950--1962, Mar. 2021.

\bibitem{Zhu2023Modeling}
L.~Zhu, W.~Ma, and R.~Zhang, ``Modeling and performance analysis for movable
antenna enabled wireless communications,'' \emph{IEEE Trans. Wireless
	Commun.}, vol.~23, no.~6, pp. 6234--6250, Jun. 2024.

\bibitem{Zhu2023Movable}
L.~Zhu, W.~Ma, B.~Ning, and R.~Zhang, ``Movable-antenna enhanced multiuser
communication via antenna position optimization,'' \emph{IEEE Trans. Wireless
	Commun.}, vol.~23, no.~7, pp. 7214--7229, Jul. 2024.

\bibitem{Ma2024MIMO}
W.~Ma, L.~Zhu, and R.~Zhang, ``{MIMO} capacity characterization for movable
antenna systems,'' \emph{IEEE Trans. Wireless Commun.}, vol.~23, no.~4, pp.
3392--3407, Apr. 2024.

\bibitem{Skouroumounis2023Fluid}
C.~Skouroumounis and I.~Krikidis, ``Fluid antenna with linear mmse channel
estimation for large-scale cellular networks,'' \emph{IEEE Trans. Commun.},
vol.~71, no.~2, pp. 1112--1125, Feb. 2023.

\bibitem{Ma2023Compressed}
W.~Ma, L.~Zhu, and R.~Zhang, ``Compressed sensing based channel estimation for
movable antenna communications,'' \emph{IEEE Commun. Lett.}, vol.~27, no.~10,
pp. 2747--2751, Oct. 2023.

\bibitem{Xiao2024Channel}
Z.~Xiao, S.~Cao, L.~Zhu, Y.~Liu, B.~Ning, X.-G. Xia, and R.~Zhang, ``Channel
estimation for movable antenna communication systems: {A} framework based on
compressed sensing,'' \emph{IEEE Trans. Wireless Commun.}, Apr. 2024, early
access, DOI: 10.1109/TWC.2024.3385110.

\bibitem{kolda2009tensor}
T.~G. Kolda and B.~W. Bader, ``{Tensor decompositions and applications},''
\emph{SIAM Rev.}, vol.~51, no.~3, pp. 455--500, Sep. 2009.

\bibitem{Zhang2022Tensor}
R.~Zhang, L.~Cheng, S.~Wang, Y.~Lou, W.~Wu, and D.~W.~K. Ng, ``{Tensor
	decomposition-based channel estimation for hybrid mmWave massive MIMO in
	high-mobility scenarios},'' \emph{IEEE Trans. Commun.}, vol.~70, no.~9, pp.
6325--6340, Sep. 2022.

\bibitem{Stegeman2007On}
A.~Stegeman and N.~D. Sidiropoulos, ``{On Kruskal’s uniqueness condition for
	the Candecomp/Parafac decomposition},'' \emph{Linear Algebra Appl.}, vol.
420, no. 2-3, pp. 540--552, Jan. 2007.

\bibitem{kay1993fundamentals}
S.~M. Kay, \emph{{Fundamentals of statistical signal processing}}.\hskip 1em
plus 0.5em minus 0.4em\relax Prentice Hall PTR, 1993.

\bibitem{Zhang2022MMV}
W.~Zhang, M.~Dong, and T.~Kim, ``{MMV-based sequential AoA and AoD estimation
	for millimeter wave MIMO channels},'' \emph{IEEE Trans. Commun.}, vol.~70,
no.~6, pp. 4063--4077, Jun. 2022.

	
\end{thebibliography}
%
\bibliographystyle{IEEEtran}

\end{document}